\documentclass[useAMS,usenatbib]{mn2e}
\usepackage{times}
\usepackage{url}
\usepackage{graphicx}
\usepackage{amssymb}
\graphicspath{{./figs/}}

\newcommand\rg{\ensuremath{r_\mathrm{g}}}
\newcommand\din{\ensuremath{\delta_\mathrm{in}}}
\newcommand\dout{\ensuremath{\delta_\mathrm{out}}}
\newcommand\rmax{\ensuremath{R_\mathrm{max}}}

\newcommand\relxill{\textsc{relxill}}

\title[Reflection fraction and black hole spin]{The role of the
  reflection fraction in constraining black hole spin}

\author[T. Dauser et al.]{T.\ Dauser$^{1}$\thanks{E-mail:
    thomas.dauser@sternwarte.uni-erlangen.de}, J. Garc\'ia$^{2}$,
  M.L.~Parker$^{3}$, A.C.~Fabian$^{3}$ and J. Wilms$^{1}$ \\
  $^{1}$ Dr.\ Karl Remeis-Observatory and Erlangen Centre for
  Astroparticle Physics,
  Sternwartstr.~7, 96049 Bamberg, Germany \\
  $^{2}$ Harvard-Smithsonian Center for Astrophysics, 60 Garden
  Street, Cambridge, MA 02138, USA \\ 
  $^{3}$ Institute of Astronomy, Madingley Road, Cambridge CB3 0HA}
\begin{document}

\pagerange{\pageref{firstpage}--\pageref{lastpage}} \pubyear{2014}

\maketitle
\label{firstpage}

\begin{abstract}
  In many active galaxies, the X-ray reflection features from the
  innermost regions of the accretion disc are relativistically
  distorted. This distortion allows us to measure parameters of the
  black hole such as its spin. The ratio in flux between the direct
  and the reflected radiation, the so-called reflection fraction, is
  determined directly from the geometry and location of primary source
  of radiation. We calculate the reflection fraction in the lamp post
  geometry in order to determine its maximal possible value for a
  given value of black hole spin. We show that high reflection
  fractions in excess of 2 are only possible for rapidly rotating
  black holes, suggesting that the high spin sources produce the
  strongest relativistic reflection features. Using simulations we
  show that taking this constraint into account does significantly
  improve the determination of the spin values. We make software
  routines for the most popular X-ray data analysis packages available
  that incorporate these additional constraints.
\end{abstract}


\section{Introduction}
\label{sec:introduction}

Reflection of X-rays off the innermost parts of the accretion disc is
highly affected by general relativistic effects, leading to strong
broadening of the reflection features \citep[see,
  e.g.,][]{Fabian1989,Laor1991,reynolds2003a,Miller2007,Dauser2010a}.
Such broadened reflection features are found in a large number of
active galactic nuclei (AGN), such as, e.g., MCG$-$6-30-15
\citep{tanaka1995a,wilms2001a,Fabian2003b,Marinucci2014a},
1H0707$-$495 \citep{Fabian2009a,Dauser2012a}, or NGC~1365,
\citep{Risaliti2013a}, Galactic Black Holes, such as Cyg~X-1
\citep{Fabian1989,Duro2011a,Tomsick2014a}, GRS~1915$+$10
\citep{Miller2013a}, or GX~339$-$4 \citep{Miller2008a,Reis2008a}, and
neutron stars \citep[e.g., Serpens~X-1;][]{Miller2013b}. Although
measurements of high-frequency Fe-K reverberation strongly support the
reflection origin of these broadened features \citep[see,
  e.g.,][]{Kara2013a,Uttley2014a}, the spectrum alone can generally be
also described by partial covering an ionized absorber \citep[see,
  e.g., ][]{Miller2008b,Miller2009b}. However, this interpretation
will not be pursued any further in this letter, as the presented
results are only applicable if a reflection origin of the broadened
spectrum is preferred.

The origin of the primary X-ray radiation causing this reflection is
still under debate. Although a standard corona of hot electrons around
the inner parts of the accretion disc is able to explain the observed
power law spectrum
\citep[e.g.,][]{Haardt1993a,Dove1997a,Belmont2008a}, it generally
fails to explain the very steep emissivities deduced from the Fe
K$\alpha$ profiles for the inner parts of the accretion disc
\citep[see,
e.g.,][]{wilms2001a,Fabian2004a,Brenneman2006a,Ponti2010a,Brenneman2011a,Gallo2011a,Dauser2012a,Risaliti2013a}.
A primary source on the rotational axis above the black hole, however,
is able yield these steeper emissivities
\citep{Matt1991a,Martocchia1996a}. If this source is associated with
the base of a jet \citep{Markoff2004a}, besides the steep emissivity,
this so called ``lamp post'' geometry is also able to explain the
spectral connection of the X-ray and radio radiation
\citep[e.g.,][]{Markoff2005a,Maitra2009a}. See \citet{Dauser2013a} for
a more detailed discussion.

In absence of opaque material that obstructs the central emission
region, an observer will always see the sum of direct radiation from
the primary source of X-rays and the reflected component. In the
classical limit, the ratio of the flux in the reflected component to
that from the primary source is the same. This ratio, referred to as
the \textit{Reflection Fraction}, $R$, can be significantly enhanced
when the source height is reduced due to the light bending effects
near the black hole \citep{Miniutti2004a}. In addition, the dependence
of the location of the inner edge of the accretion disc from the black
hole's spin, $a$, also modifies the area of the reflector and thus the
amount of reflection. The reflection fraction therefore encodes
important information about the geometry of the reflector. While the
spin is routinely determined through spectral fitting of reflection
models, most models neglected the dependence of $R$ on $a$, leaving
$R$ as a free parameter in previous studies.\footnote{Note that
  although the \texttt{pexrav} model \citep{Magdziarz1995a} includes
  $R$ as a fitting parameter, it is only suitable to describe distant
  neutral reflection.} The \relxill\ model presented by
\citet{Garcia2014a} is the first model which allows us to directly fit
the reflection fraction of a relativistically smeared reflection
spectrum.

If the geometry and location of the primary source are known, then $R$
can be easily calculated, given the angular dependence of the emission
pattern. In practice, however, except for a rough compactness
criterion \citep[source inside $< 10\,\rg$,][]{Fabian2014a}, the exact
geometry of the illumination has not been unambiguously determined for
any source yet. In this \emph{Letter} we present calculations of the
reflection fraction $R$ predicted in a lamp post geometry for
different configurations of the source height and black hole spin. A
maximum value of $R$ is then determined for a given value of spin,
which can be used to exclude unphysical solutions of low spin and
large reflection fractions. We demonstrate that this constraint has a
dramatic effect in improving the constraints on the spin
determinations, which in turn provides insights on the nature of the
geometry of the illumination source. 

\section{The Maximum Reflection Fraction} 
\label{sec:model-refl-fract}

\subsection{Determination of the Reflection Fraction}
\begin{figure}
  \centering
  \includegraphics[width=\columnwidth]{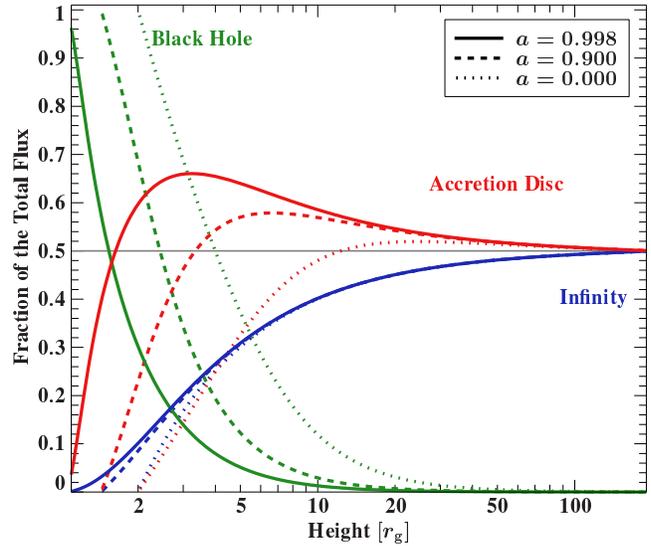}
  \caption{\label{fig:fraction} Fraction of the total flux ending up
    in the black hole (green), hitting the accretion disc (red), or
    leaving the system (blue). The calculation assumes a non-moving,
    isotropic radiating source on the axis of symmetry of the
    accretion flow, fractions are shown as a function of height of the
    primary source, given in units of the gravitational radius,
    $r_\mathrm{g} = GM/c^2$.}
\end{figure}

The reflection fraction is defined as the ratio between the reflected
and direct radiation,
\begin{equation} \label{eq:r}
  R := \frac{f_\mathrm{AD}}{f_\mathrm{INF}} = 
  \frac{\cos\din - \cos\dout}{1 + \cos\dout}
\end{equation}
where the fraction has been determined for the case of the lamp post
geometry, assuming a stationary, isotropic primary source. Here,
$f_\mathrm{AD}$ and $f_\mathrm{INF}$ denote the fraction of the
photons hitting the accretion disc or escaping to infinity,
respectively, and \din\ and \dout\ are the angles under which photons
hit the inner and outer edge of the disc, respectively
\citep{Fukumura2007a}. In the non-relativistic case and an infinitely
extended accretion disc $R=1$, i.e., half of the photons reach
infinity while the other half are reflected from the disc. This
definition of $R$ is identical to the \emph{covering fraction} used in
most reflection models such as \texttt{pexrav} \citep{Magdziarz1995a},
where it is used to describe distant neutral reflection. However, as
relativistic reflection originates from an ionized disc very close to
the black holes, the values of $R$ obtained with \texttt{pexrav}
cannot be compared with our results presented in the following. Also
note that the reflection fraction in this letter does not include any
contribution from this distant reflection by definition. In the case
of the lamp post model, as already noticed previously
\citep{Miniutti2004a}, when the source height decreases, strong
gravitational beaming reduces the number of photons that reach
infinity while $f_\mathrm{AD}$ increases, thus enhancing the
reflection features (see Fig.~\ref{fig:fraction}).  As a consequence
of these effects, the maximum possible reflection fraction depends on
the spin.

For the purpose of finding an upper limit to $R$, we use the
simplified lamp post geometry, which assumes a point-like, isotropic
source on the rotational axis of the black hole
\citep{Martocchia1996a}. First calculations of the reflection fraction
for a lamp-post like geometry were performed by \citet{Miniutti2003a}
for a ring-like primary source around an extreme Kerr black hole,
while \citet{Fukumura2007a} analysed a point-like lamp post for
different spin configurations of the black hole. We determine the
reflection spectrum using the code presented by \citet{Dauser2013a},
which is based on algorithms provided by \citet{Speith1995}. The
maximum reflection fraction is obtained for the case of the accretion
disc extending from the innermost stable circular orbit (ISCO) out to
infinity, for the cases covered by \citet{Fukumura2007a}, our results
agree with the earlier calculations.

\subsection{The Maximum Reflection Fraction}

\begin{figure}
  \centering \includegraphics[width=\columnwidth]{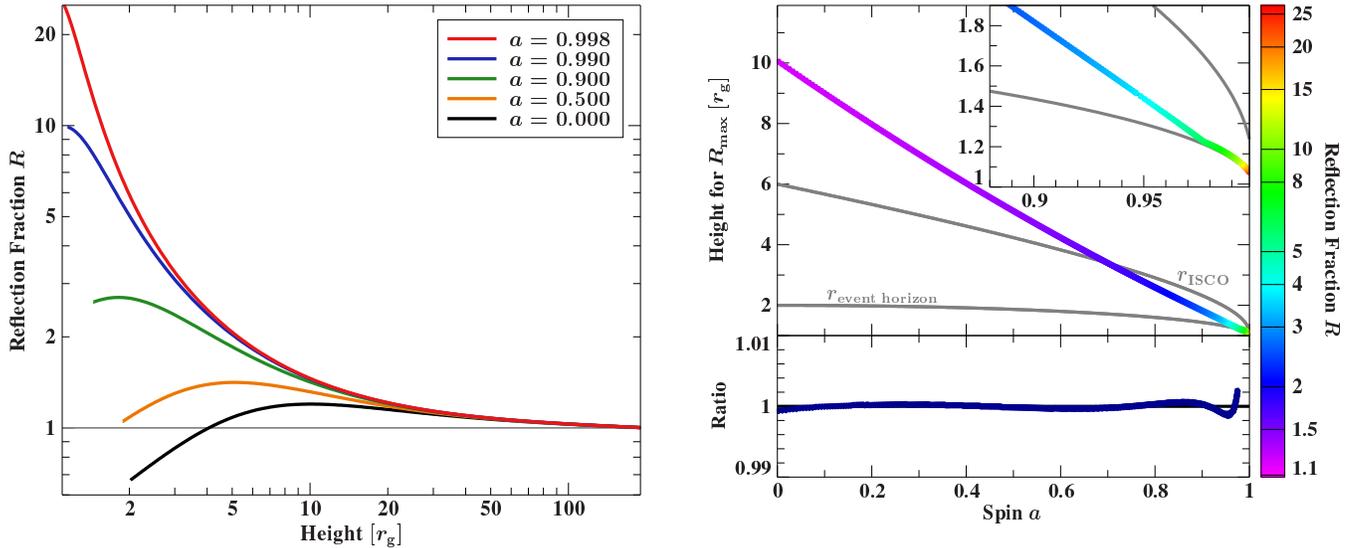}
  \caption{\label{fig:refl_frac} Reflection fraction for a point-like
    lamp post source.}
\end{figure}

\begin{figure*}
  \centering
  \includegraphics[width=\textwidth]{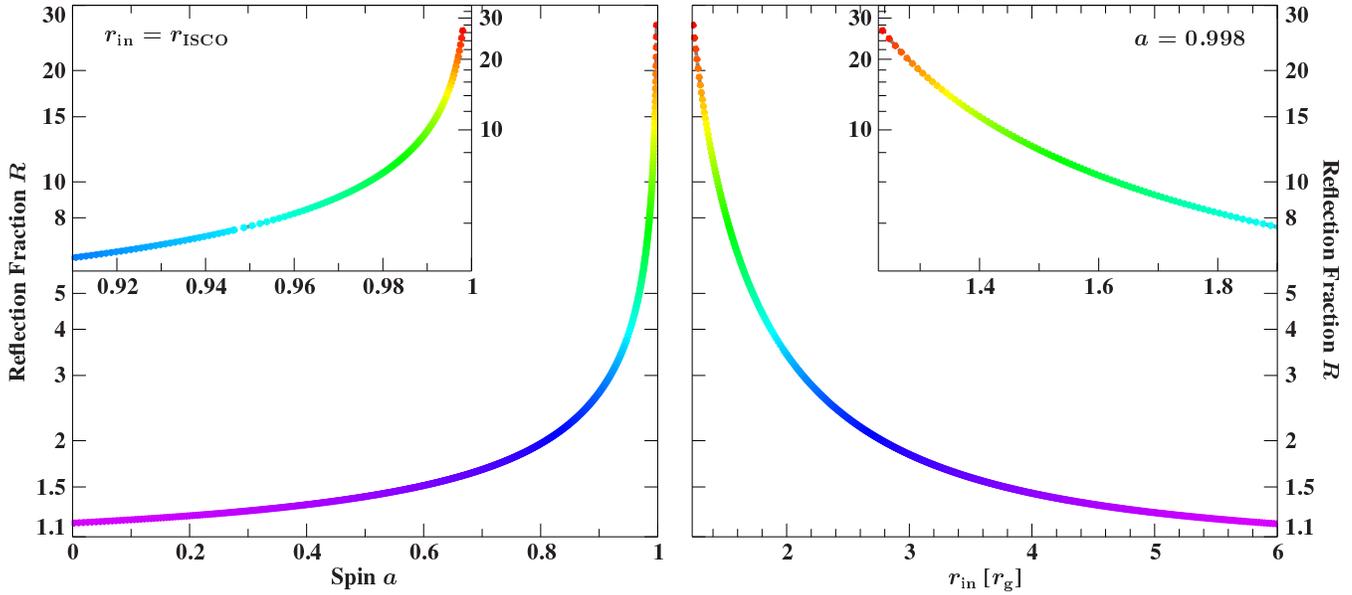}
  \caption{\label{fig:max_refl_frac} Left: maximum possible reflection
    fraction \rmax\ as a function of spin, assuming that the accretion
    disc extends down to the ISCO. Right: maximum possible reflection
    fraction for a maximally spinning black hole ($a=0.998$) as a
    function of the location of the inner edge of the accretion disc
    $r_\mathrm{in}$. The colour scale of the points is identical to
    Fig.~\ref{fig:hmax_plot}.}
\end{figure*}

Fig.~\ref{fig:refl_frac} shows how the reflection fraction varies with
the height of the source above the black hole and with black hole
spin. Except for the case of extreme spin, if the source is close to
the black hole the severe beaming of photons combined with the larger
inner disc radius reduces the flux that reaches the accretion disc and
the reflection fraction is small. On the other hand, if the primary
source is too far away from the black hole, $R\gtrsim 10\rg$,
relativistic effects diminish in importance and $R$ converges towards
its non-relativistic limit. There is therefore a spin-dependent
maximum reflection fraction, $R_\mathrm{max}$. As shown in the left
panel of Fig.~\ref{fig:max_refl_frac}, for low values of $a$, this
maximum is only slightly higher than its non-relativistic value, while
for higher spin $R_\mathrm{max}$ increases steeply.

A different way to illustrate the behaviour is shown in the right
panel of Fig.~\ref{fig:max_refl_frac}, which shows that a very similar
relation is obtained when fixing the spin, but allowing the inner edge
of the accretion disc to vary. Again, for a growing inner radius the
maximal reflection fraction quickly declines. This similarity between
changing the inner radius and changing the spin of the black is not
surprising.  As shown, e.g., by \citet{Dauser2010a}, the main effect
of the spin on the observed spectrum is to define the location of the
ISCO.

\begin{figure}
  \centering
  \includegraphics[width=\columnwidth]{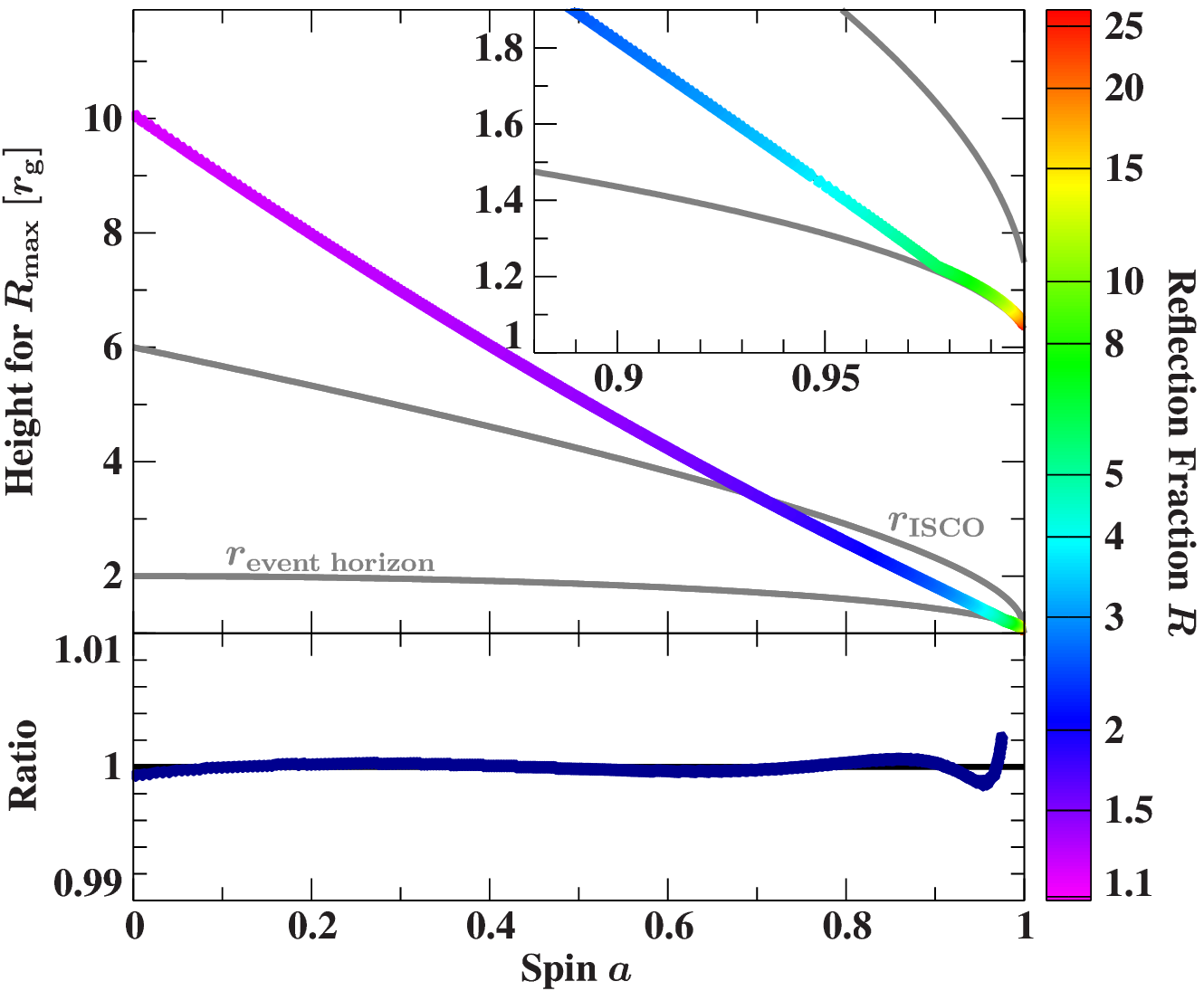}
  \caption{\label{fig:hmax_plot} The height at which the maximum
    reflection fraction \rmax\ is reached as a function of $a$ follows
    approximately a linear trend. The lower panel shows the ratio
    between the fit (Eq.~\ref{eq:fit_hmax}) and the simulated
    data. }
\end{figure}

Fig.~\ref{fig:hmax_plot} shows how the height of the point above the
black hole from which the maximum reflection is reached changes with
spin. Surprisingly, this location follows an almost linear trend which
is is well fitted by the empirical relationship
\begin{equation} \label{eq:fit_hmax}
  h_{\rmax}(a) = (1.89  a^2  - 10.86  a + 10.07)  \left( 1 +
  \frac{9.41 \times 10^{-4}}{ \log(a)} \right) 
\end{equation}
For $a < 0.975$, Eq.~\ref{eq:fit_hmax} has a precision $<0.1\%$. Above
this value of spin, the irradiating source is located directly on the
event horizon of the black hole and thus $ h(\rmax) = 1 + \sqrt{ 1 -
  a^2 }$.

We note that a point-like primary source as assumed above is only a
first-order description of the real physical system, which is likely
extended in size. However, only compact sources that are close to the
black hole will lead to a strong broadening of the Fe K$\alpha$ lines
\citep{Dauser2013a,Fabian2014a}, which is essential for reliable spin
estimates (e.g., 1H0707$-$495 or NGC~1365). Furthermore, since an
extended source can be treated as the superposition of point sources
\citep[e.g.,][]{Dauser2010a,Dauser2013a,Wilkins2012a}, it is easy to
see that the broadening of the reflection spectrum will be dominated
by the emission from the point where the flux intercepted by the disc
is at a maximum (Fig.~\ref{fig:fraction}, see also
Eq.~\ref{eq:fit_hmax} and Fig.~\ref{fig:hmax_plot} below). On the
other hand, if the primary source is off-axis and point like X-rays
will be less focused towards the disc, and therefore the maximum
reflection fraction will be less than or equal to that from a source
that is on axis. It is only in the special cases of irradiating
sources which are extended parallel to the disc surface \citep[e.g.,
  the ``disc like'' primary source inferred by][]{Wilkins2011a} or in
the case of an inflowing corona which is beamed towards the disc that
slightly higher reflection fractions than the ones presented in the
left panel of Fig.~\ref{fig:max_refl_frac} could be reached.

In some extreme cases, however, the reflection fraction measured by an
observer might be different from the theoretical one calculated
(Eq.~\ref{eq:r}). While the latter is defined as the total reflected
flux, the observer only sees the system under a certain inclination
angle and in the measurement of $R$ also needs to disentangle
inclination effects. This is especially difficult at high
inclinations, where strong beaming of the radiation along the
equatorial plane \citep[see, e.g., ][]{Sun1989a,Miniutti2004a} and
limb brightening effects are expected \citep{Garcia2014a}. Combining
these effects in a Monte Carlo simulation, \citet{Suebsuwong2006a}
have shown that larger values $R$ will only be measured for very high
inclination angles (i.e., $ \theta \gtrsim 85^\circ$). Clearly, these
inclinations are too extreme for observing the emission from the inner
regions of the accretion discs in AGN, as the reflection features
would be smeared out and the SMBH is likely to be unobservable in
presence of a torus.

\subsection{Constraints for Spin Measurements}\label{sec:constr-spin-meas}

The fact that there is a maximum reflection fraction can be used to
constrain the parameter space when modelling observational data by
excluding unphysical solutions for the spin parameter. For example,
for low spins, $a<0.5$, $\rmax<1.5$. Any solution requiring a
significantly higher reflection fraction at the same time with
requiring a low spin poses difficulties for interpreting the data. In
many cases a different combination of parameters with a high value of
the spin might be a more realistic description of the system. The
reflection fraction also puts constraints on truncated discs: An inner
radius larger than a few gravitational radii will at maximum lead to
values slightly larger than $R=1$ (Fig.~\ref{fig:max_refl_frac}, right
panel). Fits with truncated discs are therefore incompatible with
larger reflection fractions.

\begin{figure}
  \centering
  \includegraphics[width=\columnwidth]{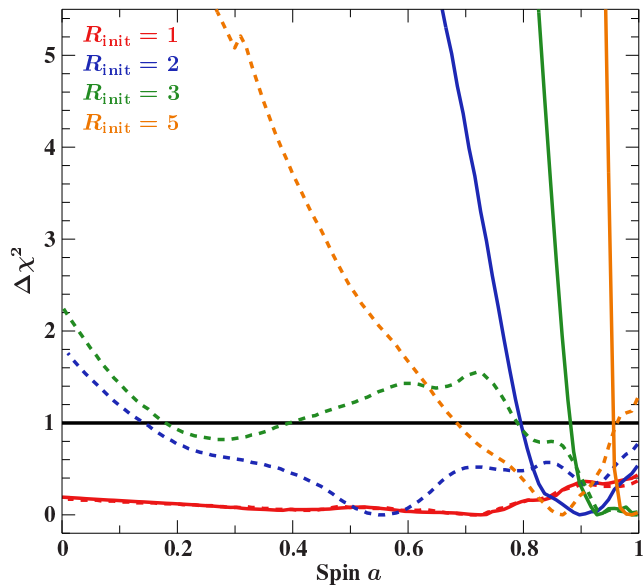}
  \caption{ Shape of the $\chi^2$-valley for a typical 100\,ks long
    AGN observation with the EPIC-pn detector on
    \textsl{XMM-Newton}. Shown is the difference of $\chi^2$ with
    respect to the best-fitting value, with $\Delta\chi^2=1$
    corresponding to the $1\sigma$ confidence interval
    \citep{Lampton1976a}. Solid lines indicate fits taking the
    reflection fraction constraint into account, dotted lines are fits
    where the constraint is ignored. The underlying continuum is based
    on the low state of 1H0707$-$495, a simple absorbed powerlaw plus
    reflection model, \textsc{tbabs$\times$relxill}, with a
    0.3--10\,keV flux of
    $10^{-12}\,\mathrm{erg}\,\mathrm{s}^{-1}\,\mathrm{cm}^{-2}$, a
    photon index $\Gamma=2$, $N_\mathrm{H}=10^{20}\,\mathrm{cm}^{-2}$,
    an emissivity index to 3, $a=0.97$, an inclination of $30^\circ$,
    and an ionization parameter $\xi =
    100\,\mathrm{erg}\,\mathrm{cm}\,\mathrm{s}^{-1}$ with an iron
    abundance that is three times solar.  In order to see how well the
    spin is constrained, the same model is fit to the simulated data
    for different values of spin $a$. All fit parameters except for
    $a$ are allowed to vary in the fits. }
  \label{fig:simulation}
\end{figure}
In order to illustrate the large potential of the presented constraint
on spin measurements, we simulate a typical observation of a low state
AGN with \textsl{XMM-Newton}. Compared to broad-band measurements
which include the reflection hump in the 20--50\,keV band and provide
a stronger constraint on $R$, measurements in the soft X-rays are
especially prone to systematic errors due to foreground emission and
absorption processes. These systematic errors tend to affect the
continuum in the iron band, and therefore the major spectral component
sensitive to $a$. Additional constraints such as that for
$R_\mathrm{max}$ are therefore especially valuable for modelling soft
X-ray data.

Fig.~\ref{fig:simulation} shows $\chi^2$-contours obtained from
fitting the continuum model to the simulated spectra, with all the
parameters free to vary, both with and without the $R_\mathrm{max}$
constraint. It is immediately evident that this constraint yields not
only a better spin estimate, but also that the unconstrained fit would
yield an incorrect value of the spin parameter ($a\sim0.55$) in the
case where $R=2$. Taking the maximum reflection fraction into account
therefore avoids possible unphysical solutions, while also reducing
the uncertainties on the spin parameter directly. Note that while a
more complicated spectrum (e.g., including distant reflection)
generally leads to weaker constraints on the spin, the improvement of
this constraint when enforcing $R_\mathrm{max}$ is still significant.

\section{Summary and Conclusions}

We presented a determination of the reflection fraction in the lamp
post geometry. Our aim was to derive a relation for the maximum
possible reflection fraction depending on the spin of the black hole.
We could show that this relation approaches the non-relativistic value
of $R=1$ quickly and therefore strongly limits the effective parameter
space of $R$ for lower values of spin. A very similar behaviour is
found for constant spin and a growing inner radius of the accretion
disc. Simulating observations and applying the additional constraint
that $R$ is at the maximum value possible for a given spin revealed
that in this way unphysical solutions can be excluded and better
constraints on $a$ could be obtained.

To allow observers to take into account these additional constraints,
we have made available a new implementation of the \relxill\
model\footnote{The \relxill\ model can be downloaded at
  \url{http://www.sternwarte.uni-erlangen.de/research/relxill/}.}
\citep{Dauser2013a,Garcia2014a} in which the reflection fraction can
either be fixed to the theoretical prediction of the lamp post model
or fitted independently.

Measurements of the reflection fractions are usually scarce and
inconsistent. However, some spectral surveys of local AGN have made
estimates of $R$ by fitting the reflection signatures
\citep[e.g.][]{nandra2007a, Brenneman2009a, Patrick2012a}. Recently,
\citet{Walton2013a} presented a study of 25 ``bare AGN'' (with no
evidence for warm absorbers or any other intervening material),
providing perhaps the first consistent estimate for the strength of
the ionized reflection (rather than the distant, neutral reflection)
in a large sample of AGN. Although their values of $R$ are
approximate, they find a strong preference for reflection fractions
near unity, which agrees well with our theoretical predictions. There
is only one case, 1H0707$-$495, for which the reflection fraction is
extremely high ($R=275$) despite its extreme spin ($a > 0.994$), but
the source was in a particularly low state when the direct power law
continuum is difficult to constrain. It is particularly interesting to
note that our simulation for $R=2$ (Fig.~\ref{fig:simulation})
resembles in many cases the $\chi^2$ confidence contours shown by
\citet[][their fig.~2]{Walton2013a}, in particular the tendency of
finding two different solutions for the spin. Good examples are
Mrk~841, 1H0323+342, UGC~6728, Mrk~359, and Mrk~1018. This behaviour
can be interpreted as a consequence of certain degeneracy between the
amount of reflection and the spin. This degeneracy can be removed by
implementing the constraint on the maximum value of $R$ presented in
this letter. Note that in certain cases measuring the reflection
component of the spectrum can be challenging, as overlying absorption
and emission processes complicate the flux determination of the
reflected radiation. Therefore the best measure is to use the flux of
the featureless Compton hump (roughly between 20 and 40\,keV). Due to
the negligible photoelectric absorption above 15keV \citep[see, e.g.,
][]{Garcia2013a}, the elemental abundances have only a negligible
influence on $R$. However, note that distant reflection might
complicate this determination.

A second observational consequence of our results is the recently
growing number of spin measurements which suggest that a large
fraction of the observed black holes are rapidly spinning \citep[see,
  e.g., ][]{Reynolds2013a}. In principle this distribution would allow
us to draw conclusions about the evolution of black hole spin
\citep[e.g., ][]{Volonteri2005a,king2008a}. However, as can be seen in
the left panel of Fig.~\ref{fig:max_refl_frac}, for not extremely
spinning black holes $R_\mathrm{max}$ is already close to 1. Sources
with lower spin are therefore expected to show a much weaker
reflection component, which, depending on the quality of the data, is
generally harder to be detected. Moreover, a low spin or a larger
source height means less pronounced relativistic reflection
\citep[e.g.,][]{Dauser2013a}.  Combined with the effect that these
configurations also lead to lower reflection fractions (see also
Fig.~\ref{fig:refl_frac}), will make the broad emission features from
such objects much harder to detect than from rapidly spinning black
holes. A stronger bias towards highly rotating black holes is
therefore expected. Also note that using standard accretion disc
theory, any flux limited sample is already biased towards detecting
high spin sources, due to a larger radiative efficiency for increasing
spin \citep[see][]{Brenneman2011a}.

\emph{Acknowledgements.} TD acknowledges support by a fellowship from
the Elitenetzwerk Bayern, and by the Deutsches Zentrum f\"ur Luft- und
Raumfahrt under contract number 50 OR 1113. JG acknowledges the
support of NASA grant NNX11AD08G. We thank John E.~Davis for the
development of the \textsc{SLxfig} module used to prepare the figures
in this letter. This research has made use of ISIS functions provided
by ECAP/Remeis observatory and MIT
(http://www.sternwarte.uni-erlangen.de/isis/).

\bibliographystyle{mn2e_williams} 
\bibliography{mnemonic,mn_abbrv,local}

\begin{thebibliography}{56}
\expandafter\ifx\csname natexlab\endcsname\relax\def\natexlab#1{#1}\fi

\bibitem[{{Belmont}, {Malzac} \& {Marcowith}(2008){Belmont}, {Malzac}, \&
  {Marcowith}}]{Belmont2008a}
{Belmont} R., {Malzac} J., {Marcowith} A., 2008, A\&A, 491, 617

\bibitem[{{Brenneman} \& {Reynolds}(2006)}]{Brenneman2006a}
{Brenneman} L.~W., {Reynolds} C.~S., 2006, ApJ, 652, 1028

\bibitem[{{Brenneman} \& {Reynolds}(2009)}]{Brenneman2009a}
{Brenneman} L.~W., {Reynolds} C.~S., 2009, ApJ, 702, 1367

\bibitem[{{Brenneman} {et~al}\mbox{.}(2011){Brenneman}, {Reynolds}, {Nowak},
  {Reis}, {Trippe}, {Fabian}, {Iwasawa}, {Lee}, {Miller}, {Mushotzky},
  {Nandra}, \& {Volonteri}}]{Brenneman2011a}
{Brenneman} L.~W. {et~al.}, 2011, ApJ, 736, 103

\bibitem[{{Dauser} {et~al}\mbox{.}(2013){Dauser}, {Garc{\'{\i}}a}, {Wilms},
  {B{\"o}ck}, {Brenneman}, {Falanga}, {Fukumura}, \& {Reynolds}}]{Dauser2013a}
{Dauser} T., {Garc{\'{\i}}a} J., {Wilms} J., {B{\"o}ck} M., {Brenneman} L.~W.,
  {Falanga} M., {Fukumura} K., {Reynolds} C.~S., 2013, MNRAS, 687

\bibitem[{{Dauser} {et~al}\mbox{.}(2012){Dauser}, {Svoboda}, {Schartel},
  {Wilms}, {Dov{\v c}iak}, {Ehle}, {Karas}, {Santos-Lle{\'o}}, \&
  {Marshall}}]{Dauser2012a}
{Dauser} T. {et~al.}, 2012, MNRAS, 422, 1914

\bibitem[{{Dauser} {et~al}\mbox{.}(2010){Dauser}, {Wilms}, {Reynolds}, \&
  {Brenneman}}]{Dauser2010a}
{Dauser} T., {Wilms} J., {Reynolds} C.~S., {Brenneman} L.~W., 2010, MNRAS, 409,
  1534

\bibitem[{{Dove} {et~al}\mbox{.}(1997){Dove}, {Wilms}, {Maisack}, \&
  {Begelman}}]{Dove1997a}
{Dove} J.~B., {Wilms} J., {Maisack} M., {Begelman} M.~C., 1997, ApJ, 487, 759

\bibitem[{{Duro} {et~al}\mbox{.}(2011){Duro}, {Dauser}, {Wilms}, {Pottschmidt},
  {Nowak}, {Fritz}, {Kendziorra}, {Kirsch}, {Reynolds}, \&
  {Staubert}}]{Duro2011a}
{Duro} R. {et~al.}, 2011, A\&A, 533, L3

\bibitem[{{Fabian} {et~al}\mbox{.}(2004){Fabian}, {Miniutti}, {Gallo},
  {Boller}, {Tanaka}, {Vaughan}, \& {Ross}}]{Fabian2004a}
{Fabian} A.~C., {Miniutti} G., {Gallo} L., {Boller} T., {Tanaka} Y., {Vaughan}
  S., {Ross} R.~R., 2004, MNRAS, 353, 1071

\bibitem[{{Fabian} {et~al}\mbox{.}(2014){Fabian}, {Parker}, {Wilkins},
  {Miller}, {Kara}, {Reynolds}, \& {Dauser}}]{Fabian2014a}
{Fabian} A.~C., {Parker} M.~L., {Wilkins} D.~R., {Miller} J.~M., {Kara} E.,
  {Reynolds} C.~S., {Dauser} T., 2014, MNRAS, 439, 2307

\bibitem[{{Fabian} {et~al}\mbox{.}(1989){Fabian}, {Rees}, {Stella}, \&
  {White}}]{Fabian1989}
{Fabian} A.~C., {Rees} M.~J., {Stella} L., {White} N.~E., 1989, MNRAS, 238, 729

\bibitem[{{Fabian} \& {Vaughan}(2003)}]{Fabian2003b}
{Fabian} A.~C., {Vaughan} S., 2003, MNRAS, 340, L28

\bibitem[{{Fabian} {et~al}\mbox{.}(2009){Fabian}, {Zoghbi}, {Ross}, {Uttley},
  {Gallo}, {Brandt}, {Blustin}, {Boller}, {Caballero-Garcia}, {Larsson},
  {Miller}, {Miniutti}, {Ponti}, {Reis}, {Reynolds}, {Tanaka}, \&
  {Young}}]{Fabian2009a}
{Fabian} A.~C. {et~al.}, 2009, Nat, 459, 540

\bibitem[{{Fukumura} \& {Kazanas}(2007)}]{Fukumura2007a}
{Fukumura} K., {Kazanas} D., 2007, ApJ, 664, 14

\bibitem[{{Gallo} {et~al}\mbox{.}(2011){Gallo}, {Miniutti}, {Miller},
  {Brenneman}, {Fabian}, {Guainazzi}, \& {Reynolds}}]{Gallo2011a}
{Gallo} L.~C., {Miniutti} G., {Miller} J.~M., {Brenneman} L.~W., {Fabian}
  A.~C., {Guainazzi} M., {Reynolds} C.~S., 2011, MNRAS, 411, 607

\bibitem[{{Garc{\'{\i}}a} {et~al}\mbox{.}(2014){Garc{\'{\i}}a}, {Dauser},
  {Lohfink}, {Kallman}, {Steiner}, {McClintock}, {Brenneman}, {Wilms},
  {Eikmann}, {Reynolds}, \& {Tombesi}}]{Garcia2014a}
{Garc{\'{\i}}a} J. {et~al.}, 2014, ApJ, 782, 76

\bibitem[{{Garc{\'{\i}}a} {et~al}\mbox{.}(2013){Garc{\'{\i}}a}, {Dauser},
  {Reynolds}, {Kallman}, {McClintock}, {Wilms}, \& {Eikmann}}]{Garcia2013a}
{Garc{\'{\i}}a} J., {Dauser} T., {Reynolds} C.~S., {Kallman} T.~R.,
  {McClintock} J.~E., {Wilms} J., {Eikmann} W., 2013, ApJ, 768, 146

\bibitem[{{Haardt}(1993)}]{Haardt1993a}
{Haardt} F., 1993, ApJ, 413, 680

\bibitem[{{Kara} {et~al}\mbox{.}(2013){Kara}, {Fabian}, {Cackett}, {Uttley},
  {Wilkins}, \& {Zoghbi}}]{Kara2013a}
{Kara} E., {Fabian} A.~C., {Cackett} E.~M., {Uttley} P., {Wilkins} D.~R.,
  {Zoghbi} A., 2013, MNRAS, 434, 1129

\bibitem[{{King}, {Pringle} \& {Hofmann}(2008){King}, {Pringle}, \&
  {Hofmann}}]{king2008a}
{King} A.~R., {Pringle} J.~E., {Hofmann} J.~A., 2008, MNRAS, 385, 1621

\bibitem[{{Lampton}, {Margon} \& {Bowyer}(1976){Lampton}, {Margon}, \&
  {Bowyer}}]{Lampton1976a}
{Lampton} M., {Margon} B., {Bowyer} S., 1976, ApJ, 208, 177

\bibitem[{{Laor}(1991)}]{Laor1991}
{Laor} A., 1991, ApJ, 376, 90

\bibitem[{{Magdziarz} \& {Zdziarski}(1995)}]{Magdziarz1995a}
{Magdziarz} P., {Zdziarski} A.~A., 1995, MNRAS, 273, 837

\bibitem[{{Maitra} {et~al}\mbox{.}(2009){Maitra}, {Markoff}, {Brocksopp},
  {Noble}, {Nowak}, \& {Wilms}}]{Maitra2009a}
{Maitra} D., {Markoff} S., {Brocksopp} C., {Noble} M., {Nowak} M., {Wilms} J.,
  2009, MNRAS, 398, 1638

\bibitem[{{Marinucci} {et~al}\mbox{.}(2014){Marinucci}, {Matt}, {Miniutti},
  {Guainazzi}, {Parker}, {Brenneman}, {Fabian}, {Kara}, {Arevalo},
  {Ballantyne}, {Boggs}, {Cappi}, {Christensen}, {Craig}, {Elvis}, {Hailey},
  {Harrison}, {Reynolds}, {Risaliti}, {Stern}, {Walton}, \&
  {Zhang}}]{Marinucci2014a}
{Marinucci} A. {et~al.}, 2014, ApJ, in press (arXiv:1404.3561)

\bibitem[{{Markoff} \& {Nowak}(2004)}]{Markoff2004a}
{Markoff} S., {Nowak} M.~A., 2004, ApJ, 609, 972

\bibitem[{{Markoff}, {Nowak} \& {Wilms}(2005){Markoff}, {Nowak}, \&
  {Wilms}}]{Markoff2005a}
{Markoff} S., {Nowak} M.~A., {Wilms} J., 2005, ApJ, 635, 1203

\bibitem[{{Martocchia} \& {Matt}(1996)}]{Martocchia1996a}
{Martocchia} A., {Matt} G., 1996, MNRAS, 282, L53

\bibitem[{{Matt}, {Perola} \& {Piro}(1991){Matt}, {Perola}, \&
  {Piro}}]{Matt1991a}
{Matt} G., {Perola} G.~C., {Piro} L., 1991, A\&A, 247, 25

\bibitem[{{Miller}(2007)}]{Miller2007}
{Miller} J.~M., 2007, ARA\&A, 45, 441

\bibitem[{{Miller} {et~al}\mbox{.}(2013{\natexlab{a}}){Miller}, {Parker},
  {Fuerst}, {Bachetti}, {Barret}, {Grefenstette}, {Tendulkar}, {Harrison},
  {Boggs}, {Chakrabarty}, {Christensen}, {Craig}, {Fabian}, {Hailey},
  {Natalucci}, {Paerels}, {Rana}, {Stern}, {Tomsick}, \& {Zhang}}]{Miller2013b}
{Miller} J.~M. {et~al.}, 2013{\natexlab{a}}, Astrophys. J., Lett., 779, L2

\bibitem[{{Miller} {et~al}\mbox{.}(2013{\natexlab{b}}){Miller}, {Parker},
  {Fuerst}, {Bachetti}, {Harrison}, {Barret}, {Boggs}, {Chakrabarty},
  {Christensen}, {Craig}, {Fabian}, {Grefenstette}, {Hailey}, {King}, {Stern},
  {Tomsick}, {Walton}, \& {Zhang}}]{Miller2013a}
{Miller} J.~M. {et~al.}, 2013{\natexlab{b}}, Astrophys. J., Lett., 775, L45

\bibitem[{Miller {et~al}\mbox{.}(2008)Miller, Reynolds, Fabian, Cackett,
  Miniutti, Raymond, Steeghs, Reis, \& Homan}]{Miller2008a}
Miller J.~M. {et~al.}, 2008, ApJ, 679, L113

\bibitem[{{Miller}, {Turner} \& {Reeves}(2008){Miller}, {Turner}, \&
  {Reeves}}]{Miller2008b}
{Miller} L., {Turner} T.~J., {Reeves} J.~N., 2008, A\&A, 483, 437

\bibitem[{{Miller}, {Turner} \& {Reeves}(2009){Miller}, {Turner}, \&
  {Reeves}}]{Miller2009b}
{Miller} L., {Turner} T.~J., {Reeves} J.~N., 2009, MNRAS, 399, L69

\bibitem[{{Miniutti} \& {Fabian}(2004)}]{Miniutti2004a}
{Miniutti} G., {Fabian} A.~C., 2004, MNRAS, 349, 1435

\bibitem[{{Miniutti} {et~al}\mbox{.}(2003){Miniutti}, {Fabian}, {Goyder}, \&
  {Lasenby}}]{Miniutti2003a}
{Miniutti} G., {Fabian} A.~C., {Goyder} R., {Lasenby} A.~N., 2003, MNRAS, 344,
  L22

\bibitem[{{Nandra} {et~al}\mbox{.}(2007){Nandra}, {O'Neill}, {George}, \&
  {Reeves}}]{nandra2007a}
{Nandra} K., {O'Neill} P.~M., {George} I.~M., {Reeves} J.~N., 2007, MNRAS, 382,
  194

\bibitem[{{Patrick} {et~al}\mbox{.}(2012){Patrick}, {Reeves}, {Porquet},
  {Markowitz}, {Braito}, \& {Lobban}}]{Patrick2012a}
{Patrick} A.~R., {Reeves} J.~N., {Porquet} D., {Markowitz} A.~G., {Braito} V.,
  {Lobban} A.~P., 2012, MNRAS, 426, 2522

\bibitem[{{Ponti} {et~al}\mbox{.}(2010){Ponti}, {Gallo}, {Fabian}, {Miniutti},
  {Zoghbi}, {Uttley}, {Ross}, {Vasudevan}, {Tanaka}, \& {Brandt}}]{Ponti2010a}
{Ponti} G. {et~al.}, 2010, MNRAS, 406, 2591

\bibitem[{{Reis} {et~al}\mbox{.}(2008){Reis}, {Fabian}, {Ross}, {Miniutti},
  {Miller}, \& {Reynolds}}]{Reis2008a}
{Reis} R.~C., {Fabian} A.~C., {Ross} R.~R., {Miniutti} G., {Miller} J.~M.,
  {Reynolds} C., 2008, MNRAS, 387, 1489

\bibitem[{{Reynolds}(2013)}]{Reynolds2013a}
{Reynolds} C.~S., 2013, Classical and Quantum Gravity, 30, 244004

\bibitem[{{Reynolds} \& {Nowak}(2003)}]{reynolds2003a}
{Reynolds} C.~S., {Nowak} M.~A., 2003, Phys. Rep., 377, 389

\bibitem[{{Risaliti} {et~al}\mbox{.}(2013){Risaliti}, {Harrison}, {Madsen},
  {Walton}, {Boggs}, {Christensen}, {Craig}, {Grefenstette}, {Hailey},
  {Nardini}, {Stern}, \& {Zhang}}]{Risaliti2013a}
{Risaliti} G. {et~al.}, 2013, Nat, 494, 449

\bibitem[{{Speith}, {Riffert} \& {Ruder}(1995){Speith}, {Riffert}, \&
  {Ruder}}]{Speith1995}
{Speith} R., {Riffert} H., {Ruder} H., 1995, Comp.\ Phys.\ Commun., 88, 109

\bibitem[{{Suebsuwong} {et~al}\mbox{.}(2006){Suebsuwong}, {Malzac}, {Jourdain},
  \& {Marcowith}}]{Suebsuwong2006a}
{Suebsuwong} T., {Malzac} J., {Jourdain} E., {Marcowith} A., 2006, A\&A, 453,
  773

\bibitem[{{Sun} \& {Malkan}(1989)}]{Sun1989a}
{Sun} W.-H., {Malkan} M.~A., 1989, ApJ, 346, 68

\bibitem[{{Tanaka} {et~al}\mbox{.}(1995){Tanaka}, {Nandra}, {Fabian}, {Inoue},
  {Otani}, {Dotani}, {Hayashida}, {Iwasawa}, {Kii}, {Kunieda}, {Makino}, \&
  {Matsuoka}}]{tanaka1995a}
{Tanaka} Y. {et~al.}, 1995, Nat, 375, 659

\bibitem[{{Tomsick} {et~al}\mbox{.}(2014){Tomsick}, {Nowak}, {Parker},
  {Miller}, {Fabian}, {Harrison}, {Bachetti}, {Barret}, {Boggs}, {Christensen},
  {Craig}, {Forster}, {F{\"u}rst}, {Grefenstette}, {Hailey}, {King}, {Madsen},
  {Natalucci}, {Pottschmidt}, {Ross}, {Stern}, {Walton}, {Wilms}, \&
  {Zhang}}]{Tomsick2014a}
{Tomsick} J.~A. {et~al.}, 2014, ApJ, 780, 78

\bibitem[{{Uttley} {et~al}\mbox{.}(2014){Uttley}, {Cackett}, {Fabian}, {Kara},
  \& {Wilkins}}]{Uttley2014a}
{Uttley} P., {Cackett} E.~M., {Fabian} A.~C., {Kara} E., {Wilkins} D.~R., 2014,
  accepted (arXiv:1405.6575)

\bibitem[{{Volonteri} {et~al}\mbox{.}(2005){Volonteri}, {Madau}, {Quataert}, \&
  {Rees}}]{Volonteri2005a}
{Volonteri} M., {Madau} P., {Quataert} E., {Rees} M.~J., 2005, ApJ, 620, 69

\bibitem[{{Walton} {et~al}\mbox{.}(2013){Walton}, {Nardini}, {Fabian}, {Gallo},
  \& {Reis}}]{Walton2013a}
{Walton} D.~J., {Nardini} E., {Fabian} A.~C., {Gallo} L.~C., {Reis} R.~C.,
  2013, MNRAS, 428, 2901

\bibitem[{{Wilkins} \& {Fabian}(2011)}]{Wilkins2011a}
{Wilkins} D.~R., {Fabian} A.~C., 2011, MNRAS, 414, 1269

\bibitem[{{Wilkins} \& {Fabian}(2012)}]{Wilkins2012a}
{Wilkins} D.~R., {Fabian} A.~C., 2012, MNRAS, 424, 1284

\bibitem[{Wilms {et~al}\mbox{.}(2001)Wilms, Reynolds, Begelman, Reeves,
  Molendi, Staubert, \& Kendziorra}]{wilms2001a}
Wilms J., Reynolds C.~S., Begelman M.~C., Reeves J., Molendi S., Staubert R.,
  Kendziorra E., 2001, MNRAS, 328, L27

\end{thebibliography}

\appendix
\label{lastpage}

\end{document}